\documentstyle[11pt,newpasp,twoside,epsf]{article}
\begin{document}
\title{STIS spectroscopy of gas disks in the nuclei of nearby, radio-loud, early-type galaxies}

\author{J. Noel-Storr\altaffilmark{1}, C. M. Carollo}
\affil{Columbia University, Astronomy Department, 550 W 120th St., New York, NY 10027, USA}
\author{S. A. Baum, R. P. van der Marel, C. P. O'Dea}
\affil{Space Telescope Science Institute, 3700 San Martin Drive, Baltimore, MD 21218, USA}
\author{G. A. Verdoes Kleijn, P. T. de Zeeuw}
\affil{Leiden Observatory, Postbus 9513, 2300 RA, Leiden, The Netherlands}

\altaffiltext{1}{Visiting Graduate Student, Space Telescope Science Institute}

\begin{abstract}
We present initial results of our analysis of line emission produced in gas disks found at the centers of a sample of nearby, radio galaxies with radio jets. We obtained data using STIS (The Space Telescope Imaging Spectrograph) at three parallel slit positions on the nucleus of each galaxy. This allows us to map the H$\alpha$ + [NII] flux, the gas radial velocity and the velocity dispersion. We find evidence of rotating disks in 11 of the sample galaxies and we can not currently rule out a rotating disk model for the remaining eight. For rotating systems, we find that the minimum central enclosed mass is greater than or similar to the predicted black hole mass based on ground-based stellar velocity dispersions. By modeling the gas dynamics we will go on to constrain the masses of the black holes. We will also investigate the properties of the gas disks themselves, giving us an insight into fueling, ionization mechanisms and the structure of the central regions.
\end{abstract}

\section{Introduction}
In seeking to understand the nature and causes of activity in galaxy nuclei, we are conducting a multi-wavelength study of a well-defined sample of 21 radio-loud, early-type galaxies in the local universe. The sample contains all nearby ($v_{\rm r} < 7000\ \rm{km s^{-1}}$), elliptical or S0 galaxies in the UGC catalog (Nilson 1973; magnitude limit $m_B < 14\fm 6$, declination range $-5\deg < \delta < 85\deg$ and angular size $\theta_p > 1\farcm 0$) that are extended radio-loud sources (larger than 10\arcsec\ on VLA A-Array maps and brighter than 150 mJy from single dish flux measurements at 1400 MHz). All of these galaxies fall into Fanaroff \& Riley's (1974) Type-I (FR-I) radio classification (see Xu et al.\ 2000, for a description of the radio properties of our sample).

Though the black hole paradigm has become widely accepted as an essential ingredient in radio galaxies, the mechanics and time-scales of fueling and jet production are poorly understood. In unified schemes (see Urry \& Padovani 1995 for a review), which suggest the appearance of AGN depends strongly on orientation, FR-I galaxies are thought to be the unbeamed population of BL-Lac objects. Understanding the central regions of such objects on scales of tens and hundreds of parsecs will allow us to better understand and characterize these connections.

\begin{figure}
\plottwo{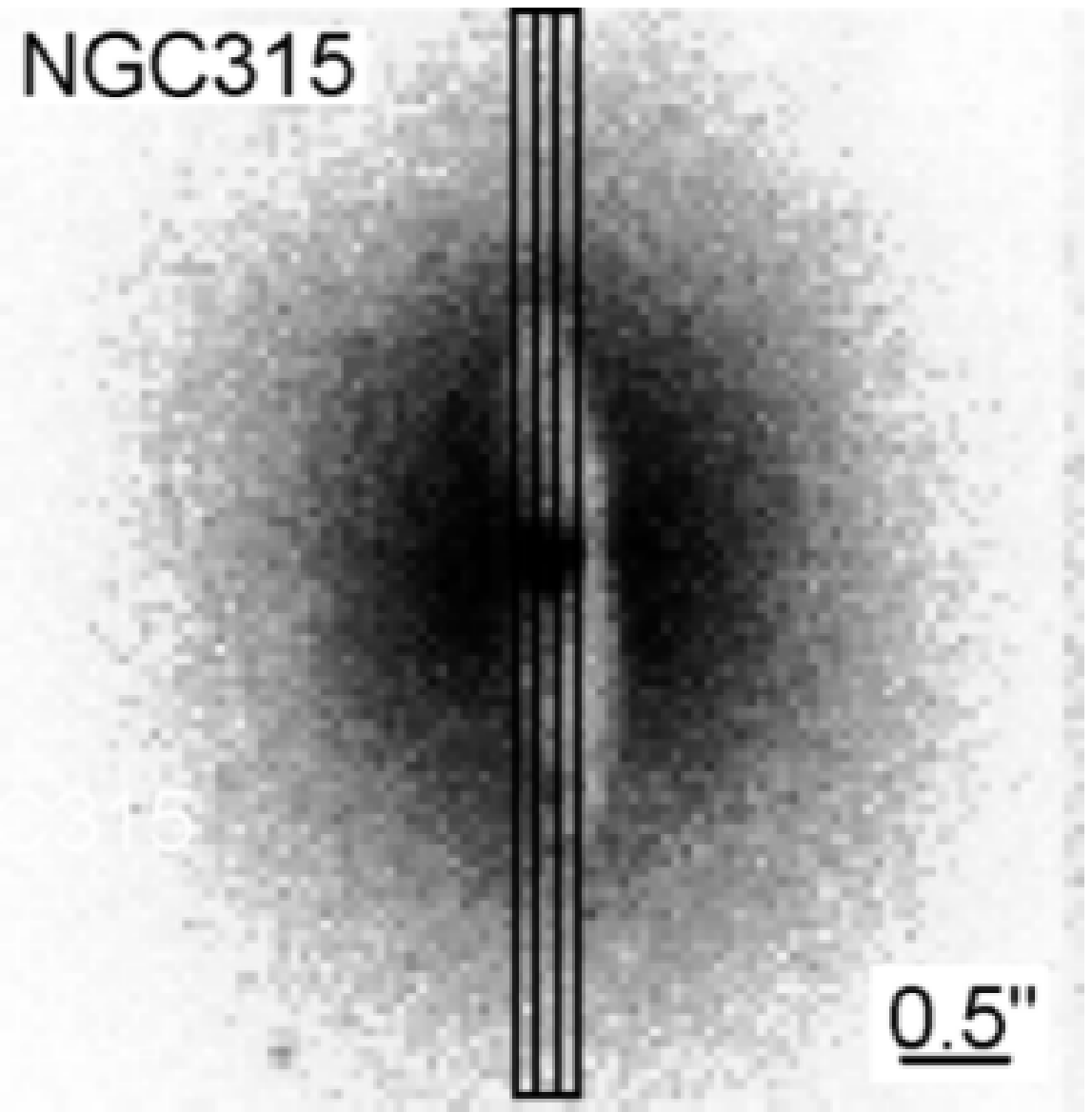}{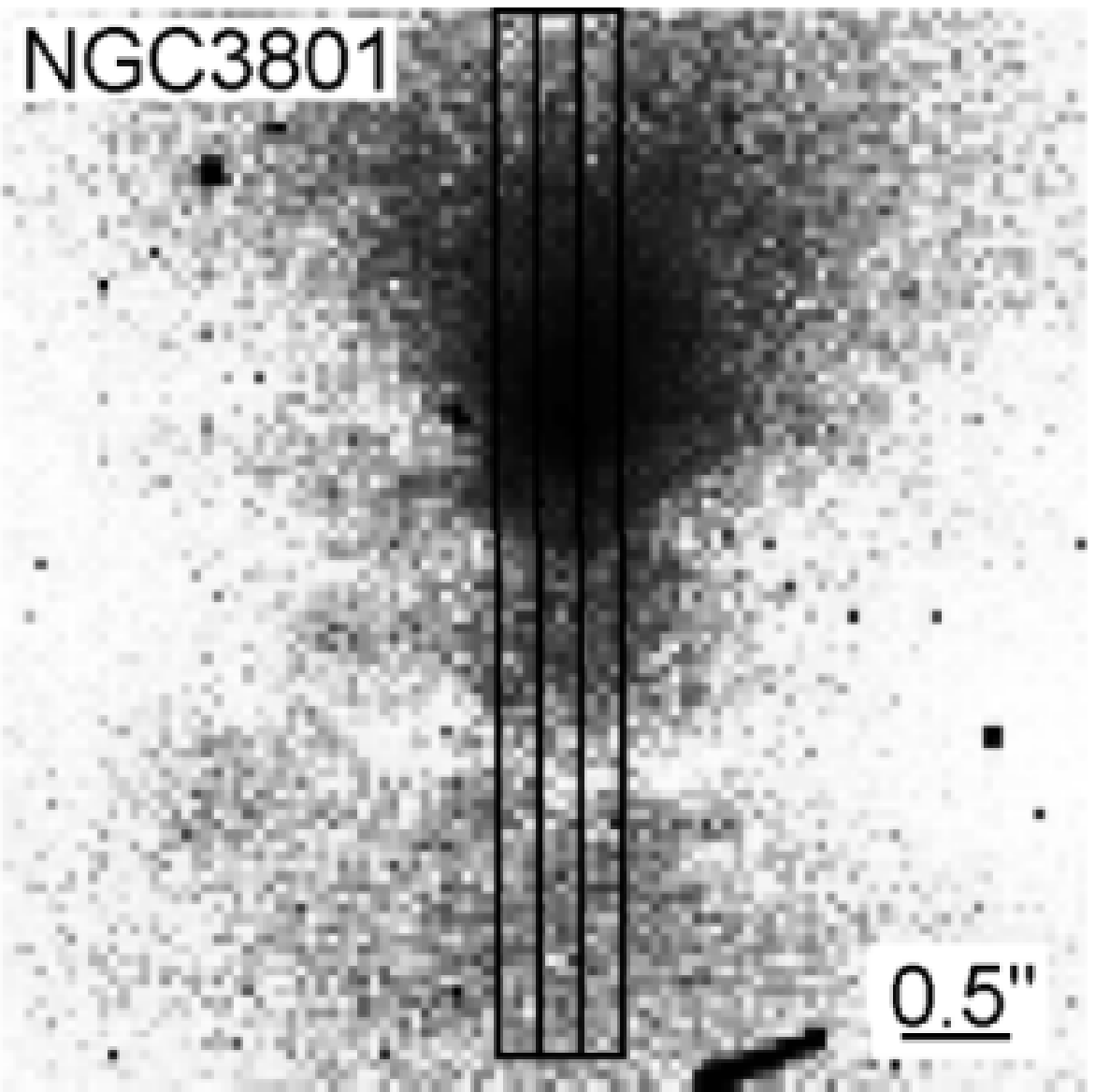}
\caption{Acquisition images showing the slit placement for two of the sample galaxies. Left: The organized center of NGC 315, with 0\farcs 1 slits. Right: the complex central morphology of NGC 3801, with 0\farcs 2 slits. Both images shown intensity as a negative grayscale.}
\end{figure}

We have observed 19 of our sample galaxies with STIS (the Space Telescope Imaging Spectrograph; see Kimble et al.\ 1998), the sample members M84 and M87 having previously been observed by others. By placing three parallel slits adjacent to each other on the galaxy nuclei (Figure 1) along the stellar major axis we have obtained sets of spectra which allow us to map, for example, the kinematics and H$\alpha$ + [NII] flux for the very central regions of each galaxy.

\section{Kinematic classifications}
By inspecting the velocity field of each galaxy it has been possible to classify them into three broad groups (see also Baum, Heckman, \& van Breugel 1992): {\em Rotators}; which show a clear, systematic, rotation pattern in their velocity field (i.e. we observe a systematic gradient in velocity across the nucleus). {\em Systematic Non-Rotators}; which show some kind of systematic behavior in their velocity field, but do not appear to be in rotation. {\em Undefined}; which do not show any clear pattern in their velocity fields.

Initially we have made use of the mean velocity dispersion ($\bar{\sigma}$) and $\Delta v = (v_{\rm max}-v_{\rm min})/2$, as estimators of the global parameters within some physical scale of the peak in emission line flux (see Table 1). 

\begin{table}
\centering
\caption{Kinematic estimators $\Delta v$ and $\bar{\sigma}$ (see text) within the given physical scales of the peak in H$\alpha$+[NII] flux for the three classes of object we define (rotators, systematic non-rotators and undefined).}
\begin{tabular}{lcccc}
\tableline
 & \multicolumn{2}{c}{r = 50 pc} & \multicolumn{2}{c}{r = 100 pc} \\
 & $\Delta v$ & $\bar{\sigma}$ & $\Delta v$ & $\bar{\sigma}$ \\
 & ${\rm (km s^{-1}})$ & $({\rm km s^{-1}})$ & ${\rm (km s^{-1})}$ & ${\rm (km s^{-1})}$\\
\tableline
Rotators (11) & $192 \pm\ 137$ & $246 \pm\ 104$ & $253 \pm\ 130$ & $211 \pm\ 72$\\
Sys. NR (3) & $100 \pm\ 19$ & $259 \pm\ 134$ & $133 \pm\ 31$ & $225 \pm\ 96$\\
Undefined (5) & $132 \pm\ 70$&$244 \pm\ 129$&$156 \pm\ 50$&$229 \pm\ 115$\\
\tableline
\tableline
\end{tabular}
\end{table}

The similarity in velocity dispersion across the categories suggests that they represent systems that are kinematically alike, and the failure to detect rotation in some cases may simply be due to adverse slit placement, the presence of dust masking part of the rotation curve or projection effects. We fail to detect rotation in galaxies that have an axis ratio of their central light distribution $b/a \ga 0.5$ (with the exception of NGC 383), i.e. the members of the sample with more nearly face-on central morphologies. Bearing this in mind, we can not rule out the possibility that all of the sample galaxies harbor gas systems of the same type viewed from a range of orientations through different obscurations.

\section{Rotating systems}
In sample members where we have been able to identify systematic rotation in the nucleus, we have made estimates of the total mass enclosed in the central region by using the maximum and minimum velocities observed (not corrected for the inclination) and the radius over which they are separated (see Table 2). 

Further modeling will allow us to improve our central mass estimates and enable us to identify and characterize the contributions of the various components that we expect, in particular the contributions of stellar populations and supermassive black holes (for example, by building on the work of van der Marel \& van den Bosch 1998; Marconi, et al.\ 2001; Sarzi, et al.\ 2001; or Barth, et al.\ 2001). This modeling will also shed light on the relative importance of non-gravitational motions in the gas.

An estimate of the anticipated black hole mass ($M_{\bullet}$), computed using the relationship found by Ferrarese \& Merritt (2001; see also Gebhardt et al.\ 2000) is provided in Table 2 ($\sigma_c$ is the central velocity dispersion corrected to an $r_e/8$ aperture). We note that all of the enclosed masses calculated (which are lower limits) are greater than or similar to the black hole mass predicted from the ground based stellar kinematics using this relation as we would expect.

\begin{table}
\caption{The mass enclosed between the maximum and minimum rotation velocities in the rotating systems (see text). For comparison the central velocity dispersion ($\sigma_{\rm c}$, corrected to a $r_{\rm e}/8$ aperture$^a$) and black hole mass ($M_{\bullet}$) inferred from the relationship of Ferrarese \& Merritt (2001) are listed. We have used $H_0 = 75\ {\rm km s^{-1} Mpc^{-1}}$.}
\centering
\begin{tabular}{cccccc}
\tableline
Galaxy & $\Delta v$ (Rot'n) & Radius & $M_{\rm Enclosed}^b$ & $\sigma_c$ &
$M_{\bullet}$\\
& $({\rm km s^{-1}})$ & $({\rm pc})$ & $(M_{\sun})$ & $({\rm km s^{-1}})$ &
$(M_{\sun})$\\
\tableline
NGC 315 & 344.8 & 25 & $7.0\times 10^8$ & 295 & $8.2\times 10^8$\\ 
NGC 383 & 420.2 & 48 & $2.0\times 10^9$ & 254 & $4.0\times 10^8$\\
NGC 741 & 530.3 & 138 & $9.1\times 10^9$ & 265 & $4.9\times 10^8$\\
UGC 7115 & 413.3 & 44 & $1.8\times 10^9$ & 175 & $6.9\times 10^7$\\
NGC 4261$^c$ & 174.0 & 73 & $5.1\times 10^8$ & 291 & $7.6\times 10^8$ \\
NGC 4335 & 305.9 & 121 & $2.6\times 10^9$ &\multicolumn{2}{c}{Not available} \\
NGC 5127 & 315.3 & 190 & $4.4\times 10^9$ & 178 & $7.5\times 10^7$\\
NGC 5141 & 471.9 & 87 & $4.5\times 10^9$ &\multicolumn{2}{c}{Not available} \\
NGC 7052$^d$ & 531.5 & 54 & $3.6\times 10^9$ & 247 & $3.5\times 10^8$ \\
UGC 12064 & 229.1 & 34 & $4.1\times 10^8$ & 257 & $4.2\times 10^8$\\
NGC 7626 & 472.6 & 34 & $1.8\times 10^9$ & 248 & $3.6\times 10^8$\\
\tableline
\tableline
\end{tabular}
\par\small{$^a$ $\sigma_{\rm c}$ from the LEDA database; correction applied following Jorgensen et al.\ (1995)}
\par\small{$^b$ $M_{\rm Enclosed}$ is the minimum enclosed mass based on the quoted $\Delta v$ and radius without correction for inclination angle.}
\par\small{$^c$ NGC 4261: $M_{\bullet}$ measured to be $4.9\times 10^8 M_{\sun}$ (Ferrarese, Ford \& Jaffe 1996)}
\par\small{$^d$ NGC 7052: $M_{\bullet}$ measured to be $3.3\times 10^8 M_{\sun}$ (van der Marel \& van den Bosch 1998)}
\end{table}


\begin{references}
\reference Barth, A. J., et al.\ 2001, \apj, in press (astro-ph/0012213)
\reference Baum, S. A., Heckman, T. M., \& van Breugel, W. 1992, \apj, 389, 208
\reference Fanaroff, B. L., \& Riley, J. M. 1974, \mnras, 167, 31P
\reference Ferrarese, L., Ford, H. C., \& Jaffe, W. 1996, \apj, 470, 444
\reference Ferrarese, L., \& Merritt, D. 2001, \apj, 547, 140
\reference Gebhardt, K., et al.\ 2000, \apj, 539, 13
\reference Jorgensen, I., Franx, M., \& Kjaergaard, P. 1995, \mnras, 276, 1341
\reference Kimble, R., et al.\ 1998, \apj, 492L, 83
\reference Marconi, A., et al.\ 2001, \apj, 549, 915
\reference Nilson, P. 1973 The Uppsala General Catalog of Galaxies [UGC], (Uppsala: Astronomiska Observatorium)
\reference Sarzi, M., et al.\ 2001, \apj, 550, 65
\reference Urry, C. M., \& Padovani P. 1995, \pasp, 107, 803
\reference van der Marel, R. P., \& van den Bosch, F. C. 1998, \aj, 116, 2220
\reference Xu, C., Baum, S. A., O'Dea, C. P., Wrobel, J. M., \& Condon, J. J. 2000, \aj, 120, 2950
\end{references}
\end{document}